\documentclass[pra,twocolumn,superscriptaddress,nofootinbib]{revtex4}
\usepackage{amsmath}
\usepackage{amsfonts}
\usepackage{amssymb}
\usepackage{graphicx}

\begin{document}

\title{Degradation of a quantum directional reference frame as a random walk}
\author{Stephen D. Bartlett}
\affiliation{School of Physics, The University of Sydney, Sydney,
    New South Wales 2006, Australia}
\author{Terry Rudolph}
\affiliation{Optics Section, Blackett Laboratory, Imperial College
    London, London SW7 2BZ, United Kingdom}%
\affiliation{Institute for Mathematical Sciences, Imperial College
    London, London SW7 2BW, United Kingdom}
\author{Barry C. Sanders}
\affiliation{Institute for Quantum Information Science, University
    of Calgary, Alberta T2N 1N4, Canada}%
\affiliation{Centre of Excellence for Quantum Computer Technology,
    Macquarie University, Sydney, New South Wales 2109, Australia}
\author{Peter S. Turner}
\affiliation{Institute for Quantum Information Science, University
    of Calgary, Alberta T2N 1N4, Canada}
\date{31 January 2007}

\begin{abstract}
We investigate whether the degradation of a quantum directional reference frame
through repeated use can be modelled as a classical direction undergoing a
random walk on a sphere.  We demonstrate that the behaviour of the
\emph{fidelity} for a degrading quantum directional reference frame, defined
as the average probability of correctly determining the orientation of a
test system, can be fit precisely using such a model.  Physically, the
mechanism for the random walk is the uncontrollable back-action on the
reference frame due to its use in a measurement of the direction of another
system.  However, we find that the magnitude of the step size of this random
walk is not given by our classical model and must be determined from the
full quantum description.
\end{abstract}

\maketitle

\section{Introduction}

Reference frames are physical objects subject to the laws of physics
like any other.  Although it is common, especially within quantum
physics, to treat reference frames as non-dynamical objects outside
of the theory, there are a wide variety of situations where this
assumption is not valid.  Reference frames are quantifiable
resources~\cite{BRS06}: they are non-trivial to prepare and to
share~\cite{Per01}, their imperfections can result in deleterious
consequences for performing high-precision
measurements~\cite{Enk02,Gea02,Tyc04}, and they can \emph{degrade}
with repeated use~\cite{BRST,Pou06}.

The issue of degradation of a reference frame is particularly
important for quantum computation, where a large number of
high-precision measurements must be performed.  In many
implementations, such measurements are performed relative to a
reference frame that is described by a finite quantum system; for
example, a small-amplitude coherent state of the electromagnetic
field~\cite{Enk02,Gea02}, the proposed single-spin measurement
technique using magnetic resonance force microscopy~\cite{Rug04}, or
the single-electron transistors used for measurement of
superconducting qubits~\cite{Mak01}.  When measuring an ensemble of
systems for which the mean polarization is precisely zero, the
number of measurements for which a quantum reference frame for
direction or for phase can be used has been shown to scale
quadratically rather than linearly in the size of the reference
frame~\cite{BRST}.  This result provides promise for the use of
microscopic or mesoscopic reference frames in performing repeated
high-precision measurements where degradation of the reference frame
is significant, provided that quantum information is encoded in
states with zero mean polarization.

In~\cite{Pou06}, it was argued that this quadratic scaling behaviour
is the result of fluctuations in the polarization of the measured
systems.  A primary motivation for the current paper was to address
the question of whether the quadratic scaling behaviour was a
manifestation of the well known quadratic behaviour of a classical
random walk's mean displacement. In this paper we answer this
question in the negative. Although a classical random walk does
capture the functional form of the degradation process, the
\textit{step size} of the walk (which is what ultimately determines
the quadratic scaling of the longevity described above) must be
determined from the full quantum description. Once the step size is
fit to the quantum model, however, the classical random walk
determines at least one measure of the frame's degradation
perfectly.

The degradation of a quantum directional reference frame (DRF) can
be quantified by the decrease in measurement fidelity (the
probability of correctly measuring a spin that is known to be
aligned/anti-aligned with the frame) as a function of the number of
previous measurements for which it has been used. Although the
mathematical framework for calculating the decrease in measurement
fidelity has been established~\cite{BRST}, the simple connection
with a random walk shown here provides two important advantages. One
advantage is that the random walk is easier to model so the
incorporation of reference frame degradation into quantum computing
simulations will be simplified by this random walk analogy rather
than having to account for a large Hilbert space. The second
advantage is that the random walk analogy provides a simple and
elegant conceptual framework for the the effects on reference frames
by sequential measurements, which will make considerations of
general problems involving reference frame degradation easier to comprehend
and solve.

The structure of the paper is as follows.  In Sec.~\ref{eq:QReview}, we 
review the results of the degradation of a
quantum DRF~\cite{BRST}.  In Sec.~\ref{sec:Semiclassical}, we
present a semi-classical model for a DRF. We demonstrate how a
classical model with some uncertainty in its direction can be
characterized by a probability distribution on a sphere.  We
initiate our semi-classical DRF in a probability distribution that
is comparable to the quantum DRF in~\cite{BRST}. At this point, in
order to introduce a degradation due to its use in measurement, we
construct a semi-classical measurement theory for the DRF based on
two simple assumptions.  We argue that this measurement theory leads
to a spherical random walk on the probability distribution of this
classical DRF.  Finally, we show that the \emph{form} of the
degradation exactly matches that of quantum case found
in~\cite{BRST}, and that a single fitted parameter -- the classical
step size -- yields a \emph{value} of the classical degradation that
exactly matches the quantum. Conclusions are presented in
Sec.~\ref{sec:Conclusions}.

\section{Degradation of a quantum DRF}
\label{eq:QReview}

In this section, we review the results of~\cite{BRST}, which
characterize the degradation of a quantum DRF.  The key result we
will reproduce with a semi-classical model is the rate of decrease
of the measurement fidelity in terms of the number of previous
measurements for which the reference frame was used.

We use a spin-$j$ system for our quantum DRF, with Hilbert space
$\mathbb{H}_{j}$.  We choose the initial quantum state of the
spin-$j$ system to be $\rho^{(0)} = |j,j\rangle\langle j,j|$, which
was determined in~\cite{BRST} to be the initial state that maximizes
the initial success probability. (This quantum reference frame is
aligned in the $+z$ direction relative to a background DRF, which is
essentially a choice of gauge.)

The systems to be measured against the quantum DRF will be spin-1/2
systems, each with a Hilbert space $\mathbb{H}_{1/2}$. We consider
the situation wherein the initial state of each such system is the
completely mixed state $I/2$, where $I$ is the identity operator on
$\mathbb{H}_{1/2}$.  The use of maximally mixed qubits corresponds
to having no information about the state of the spin-1/2 systems
prior to measurement.  In particular, this condition ensures that
the qubits have zero mean polarization with respect to the quantum
DRF, i.e., the qubits and the quantum DRF are initially completely
uncorrelated.  (Note that, if the measured systems have a non-zero
mean polarization with respect to the DRF, even a very small one,
the resulting measurements will induce a drift on the DRF resulting
in a linear scaling~\cite{Pou06}.)

Our quantum DRF will be used to measure many such independent
spin-1/2 systems sequentially. We shall assume trivial dynamics
between measurements, and thus our time index will simply be an
integer specifying the number of measurements that have taken place.
The state of the DRF following the $n^\text{th}$ measurement is
denoted $\rho^{(n)}$, with $\rho^{(0)}$ denoting the initial state
of the DRF prior to any measurement.

A measurement of the relative orientation of a spin-1/2 particle to
a spin-$j$ system is represented by operators that are invariant
under collective rotations. The measurement that provides the
maximum information gain about the relative orientation between a
spin-$j$ and a spin-$1/2$ system has been determined
in~\cite{BRS04b} to be a measurement of the magnitude of the total
angular momentum $\hat{J}^{2}$. That is, it is the two-outcome
projective measurement $\{\Pi _{+}\equiv \Pi _{j+1/2},\Pi _{-}\equiv
\Pi _{j-1/2}\}$ on $\mathbb{H}_{j}\otimes \mathbb{H}_{1/2}$.

The evolution of the DRF will depend on the measurement
record.  In general, even if one keeps the entire measurement record
and determines the conditional state of the DRF based on this
record, there will be a degradation of the DRF (specifically, a
non-unitary evolution of the reduced density matrix describing the
DRF) due to entanglement between the state of the DRF and the
measured spins.  The magnitude of this degradation will depend
explicitly on the measurement record.  In order to quantify the
\emph{expected} (or average) degradation, we consider averaging over
all possible measurement records.  Note that this averaging is
equivalent to \emph{discarding} the measurement record, and we consider
this expected degradation in what follows.  We note that an
interesting avenue for future work would be to address the
worst-case scenario as opposed to the average case.

The evolution of the quantum DRF as a result of the $n^\text{th}$
measurement is
\begin{equation}
  \rho ^{(n+1)}=\mathcal{E}(\rho ^{(n)})  \label{eq:DecohMap}
\end{equation}
where the superoperator $\mathcal{E}$ is given by
\begin{equation}
  \mathcal{E}(\rho)=\text{Tr}_{S}\Bigl(\sum_{c\in \{+,-\}}\Pi _{c}(\rho
  \otimes I/2)\Pi_{c}\Bigr)\,,  \label{eq:DecohMap2}
\end{equation}
with $\text{Tr}_{S}$ the partial trace over $\mathbb{H}_{1/2}$.

The map $\mathcal{E}$ can be written using the operator-sum
representation~\cite{Nie00} as
\begin{equation}
  \mathcal{E}(\rho )=\frac{1}{2}\sum_{\stackrel{c\in \{+,-\}}{a,b\in
  \{0,1\}}}E_{ab}^{c}\rho E_{ab}^{c\dag}\,,
\label{eq:RFdecoherence}
\end{equation}
where $E_{ab}^{c}\equiv \langle a|\Pi _{c}|b\rangle$ is a Kraus
operator on $\mathbb{H}_{j}$ and $\{|0\rangle,|1\rangle \}$ is a
basis for $\mathbb{H}_{1/2}$. These operators can be
straightforwardly determined in terms of Clebsch-Gordon
coefficients.  The resulting evolution has not been solved
analytically; a closed-form solution for the limiting case $j\gg 1$
is presented in Ref.~\cite{BRST}.

We quantify the quality of a quantum DRF as the average probability
of a successful estimation of the orientation of a fictional
``test'' spin-1/2 system which is, with equal probability, either
aligned or anti-aligned with the background $+z$-axis.  Denote the
pure state of the test spin-1/2 system that is aligned
(anti-aligned) with the initial DRF by $|0\rangle$ ($|1\rangle$).
For a spin-1/2 system prepared in the state $|0\rangle$ or
$|1\rangle$ with equal probability, the average probability of
success using a quantum DRF state $\rho$ is
\begin{equation}
  F_Q =\frac{1}{2}\mathrm{Tr}_{R}(\rho(E_{00}^{+}+E_{11}^{-}))\,,
\end{equation}
where the subscript $Q$ denotes that this is a fully quantum result.
We denote this quantity as the \emph{quantum average measurement
fidelity}.

Although Eq.~\eqref{eq:RFdecoherence} has not been solved
analytically, it is still possible to obtain an \emph{exact}
expression for the evolution of the quantum average measurement
fidelity under this evolution.  In~\cite{BRST}, it is found that
the solution for $\rho^{(n)}$, given the initial state $\rho^{(0)} =
|j,j\rangle\langle j,j|$, yields a quantum average measurement
fidelity $F_Q(n)$ that decreases with $n$ as
\begin{equation}
  F_Q(n)=\frac{1}{2}+\frac{j}{2j+1}\Bigl(1-\frac{2}{(2j+1)^{2}}
  \Bigr)^{n}\,.
  \label{eq:PsuccessSU(2)CS}
\end{equation}
This expression, which we emphasize is exact, implies that
the number of measurements for which an DRF is useful increases
\emph{quadratically} with $j$, the size of the reference frame.

\section{A semi-classical model}
\label{sec:Semiclassical}

In this section, we demonstrate that the complete quantitative
behavior of the degradation of a DRF can be modeled as an ideal DRF
undergoing a (classical) random walk on the sphere. For such a
description, we will develop a semi-classical measurement theory --
one that describes measurement with respect to an ideal reference
frame about which there is some classical uncertainty, and which
includes a \emph{back-action} due to the measurement that randomly
``kicks'' the reference frame, yielding a random walk.

\subsection{Probability distributions of a DRF}

Consider an ideal classical DRF.  (This DRF could be
realized, say, by a spin system as described above with $j\to
\infty$.)  The possible configurations of this DRF are described by
the unit sphere, $S^2$, with coordinates $(\theta,\phi)$ given by
the polar angle $\theta\in[0,\pi]$ and the azimuthal angle
$\phi\in[0,2\pi)$. Such an ideal classical DRF allows for
projective measurements to be performed in the angular momentum
basis defined by $(\theta,\phi)$.

To model degradation of an ideal classical DRF, we consider
classical probability distributions $p(\theta,\phi)$ on the unit
sphere $S^2$. These probability distributions describe the
experimenter's knowledge about the orientation of the classical
DRF. The usual properties of probability distributions
must apply, namely normalisation and positivity:
\begin{gather}
  \label{eq:Normalized}
  \int_0^{2\pi}\frac{\mathrm{d}\phi}{2\pi}
  \int_0^\pi\frac{\sin\theta\mathrm{d}\theta}{2}\,p(\theta,\phi)=1\,,
  \\ p(\theta,\phi)\geq 0\quad \forall\;(\theta,\phi)\in S^2\,.
  \label{eq:Positive}
\end{gather}

All functions on the sphere can be expanded in terms of spherical
harmonics $Y_{\ell m}(\theta,\phi)$.  In this paper, we will only
make use of distributions that are azimuthally symmetric, i.e.,
distributions that do not depend upon the azimuthal angle $\phi$.
Such distributions can be expanded in terms of the spherical
harmonics with $m=0$.  Because $Y_{\ell 0}$ is proportional to the
Legendre polynomial $P_\ell(\cos\theta)$, we have
\begin{equation}
  \label{eq:DistnSphericalHarmonics}
  p(\theta) = \sum_{\ell=0}^\infty\,c_\ell P_\ell(\cos\theta)\,.
\end{equation}
Two Legendre polynomials that will play a major r\^ole in what
follows are $P_0(x)=1$ and $P_1(x)=x$.  The inner product
\begin{equation}\label{eq:OrthogonalLegendre}
    \int_0^\pi\frac{\sin\theta\mathrm{d}\theta}{2}\, P_\ell(\cos\theta)
    P_{\ell'}(\cos\theta) = \frac{1}{2\ell +1} \delta_{\ell\ell'}
    \,,
\end{equation}
will also be useful.

\subsection{An initial distribution comparable with a finite quantum DRF}

We will initiate our classical DRF (at $n=0$, prior to the
first measurement) aligned in the $+z$ direction, just as in the
fully quantum description.  (Again, this is essentially a choice of
gauge.)  Also, we choose an initial probability distribution that
has classical uncertainties comparable with those of the initial
quantum state $\rho^{(0)} = |j,j\rangle\langle j,j|$ that was used
in Sec.~\ref{eq:QReview}. Noting that $\rho^{(0)}$ is invariant
under rotations about the $z$-axis, the initial classical
distribution should also be so, i.e., independent of $\phi$. To this
end, we choose the initial classical distribution to be
\begin{equation}
\label{eq:InitialClassicalDistribution}
    p^{(0)}(\theta) = (4j+1)\bigl[\cos(\theta/2)\bigr]^{8j} .
\end{equation}
Equation~(\ref{eq:InitialClassicalDistribution}) is a valid probability distribution, satisfying
Eqs.~\eqref{eq:Normalized} and \eqref{eq:Positive}, and which we now
argue has the correct angular uncertainties, comparable with those
of the quantum state $|j,j\rangle$.

The state $|j,j\rangle$ can be viewed as an SU(2) coherent
state~\cite{Are72,PER} and saturates the angular-momentum
uncertainty relations $(\Delta J_x)^2 (\Delta J_y)^2 \geq
\frac{1}{4}\langle J_z \rangle^2$, with $(\Delta J_x)^2 = (\Delta
J_y)^2 = j/2$.  The angular uncertainty (variance) of the state
$|j,j\rangle$, then, is $(\Delta \theta)^2 = (\Delta J_x)^2/\langle
J_z \rangle^2 = 1/(2j)$.

The distribution $p(\theta)$ possesses the same angular variance, as
follows.  For small $\theta$ and large $j$, we can use the
approximation~\cite{RdGS01}
\begin{equation}\label{eq:ApproxGaussian}
  \bigl[\cos(\theta/2)\bigr]^{8j} \rightarrow \exp(-j\theta^2) \,.
\end{equation}
Viewed as a Gaussian distribution for small $\theta$, this has
variance $\sigma^2 = 1/(2j)$.  Thus, the probability distribution of
Eq.~\eqref{eq:InitialClassicalDistribution} is (i) aligned along the
$+z$-direction, (ii) azimuthally symmetric, and (iii) has angular
uncertainties that are equal to those of the quantum state
$|j,j\rangle$.

\subsection{Semi-classical measurement theory}

We now introduce a semi-classical measurement theory that allows us
to calculate the probability of correctly measuring a \emph{known}
spin, either aligned or anti-aligned with the background $z$-axis,
using an ideal classical DRF with an uncertainty in its
direction (i.e., described by a probability distribution).

\subsubsection{Measurement fidelity}

The measurement of a single spin-1/2 particle relative to an ideal
classical DRF aligned along the $+z$-axis can be described
as a projective measurement of the qubit in the $\{ |0\rangle,
|1\rangle\}$ basis. If this ideal classical DRF is
\emph{misaligned} by a known angle $\theta$ relative to the
background $+z$-axis, the measurement of a spin-1/2 particle
relative to this misaligned DRF is also described by a projective
measurement but now in the basis $\{ (\cos(\theta/2)|0\rangle +
\sin(\theta/2)|1\rangle), (\cos(\theta/2)|1\rangle -
\sin(\theta/2)|0\rangle) \}$.  Consider a spin-1/2 particle that is
known to be aligned with the background $+z$-axis. The probability
that a measurement of this particle relative to the misaligned frame
will yield the correct result is $P = (1+\cos\theta)/2 =
\cos^2\theta/2$. Thus, a misalignment of an ideal classical
DRF leads to a finite probability of error when measuring the
direction of objects that are known to be aligned (or anti-aligned)
to the $z$-axis.

Now consider extending this measurement theory to an ideal
classical DRF with a distribution of misalignments $p(\theta)$
compared to the background $z$-axis.  The probability of correctly
measuring the direction of a spin-1/2 particle that is known to be
aligned or anti-aligned to the $z$-axis is given by
\begin{equation}\label{eq:cfid}
  F_C = \int_0^\pi \frac{\sin\theta\mathrm{d}\theta}{2} \: p(\theta) \: \cos^2(\theta/2)
  \,,
\end{equation}
which we will call the \emph{classical average measurement fidelity}.

We can expand $p(\theta)$ in the basis of Legendre polynomials as in
Eq.~\eqref{eq:DistnSphericalHarmonics}.  Because $\cos^2(\theta/2)$
can be simply expressed in terms of Legendre polynomials as
$\cos^2(\theta/2) = \frac{1}{2}(P_0(\cos\theta) + P_1(\cos\theta))$,
and because the Legendre polynomials form an orthogonal basis, the
classical average measurement fidelity depends only upon the $\ell=0$
and $\ell=1$ terms, i.e.,
\begin{align}
  F_C &= \int_0^\pi\frac{\sin\theta\mathrm{d}\theta}{2} \Bigl( \sum_{\ell=0}^\infty\,c_\ell
  P_\ell(\cos\theta) \Bigr) \cos^2(\theta/2)
  \nonumber \\
  &= \sum_{\ell=0}^\infty\,c_\ell
  \int_0^\pi \frac{\sin\theta\mathrm{d}\theta}{2} \:
  P_\ell(\cos\theta) \: \tfrac{1}{2}\Bigl(P_0(\cos\theta) + P_1(\cos\theta)\Bigr)
  \nonumber \\
  &= \tfrac{1}{2}\bigl( c_0 + \tfrac{1}{3}c_1 \bigr) \,.
  \label{eq:cfidc0c1}
\end{align}

We note that this classical average measurement fidelity $F_C$ can
be compared directly with the quantum average measurement fidelity
$F_Q$: both quantify the average probability of a successful
estimation of a known spin that is either aligned or anti-aligned
with the $+z$-axis.

\subsubsection{Back-action}

Our semi-classical measurement theory will also incorporate a form
of back-action on the reference frame, i.e., an uncontrollable
disturbance of the direction of the reference frame as the result of
a measurement.  A natural model for this back-action is as follows:
as a result of a measurement of a spin, the state of the DRF is
``kicked'' in a random direction (sampled uniformly from the
interval $[0,2\pi)$) by a fixed angular distance $\alpha$. Repeated
measurements using this DRF, then, can be modeled by a random walk
on the sphere.

We note that the step size $\alpha$ is a free parameter in our
measurement theory; it will be determined by fitting the behavior
with the fully quantum evolution.  However, it would be natural for
this step size to vary as the ratio of the size of the system being
measured to the size of the reference frame.  In the quantum
scenario that we wish to model, where $\hbar j$ is the angular
momentum of the quantum reference frame and $\hbar/2$ is the angular
momentum of the spin-$1/2$ systems being measured, this ratio scales
as $1/j$.  We will find that our fit to the quantum model agrees
with this intuition.

\subsection{Classical random walk on the sphere}

We now consider the evolution of a probability distribution $p$
undergoing a random walk on the sphere~\cite{RU}.  Let the initial
distribution be denoted $p^{(0)}$, and denote the distribution after
$n$ steps by $p^{(n)}$.  Our random walk on the sphere is defined
such that, at each step, it is equally probable to make a step of
fixed angular size $\alpha$ in any direction.  The evolution of the
probability distribution undergoing the random walk is then given by
\begin{equation}
  p^{(n+1)}(\theta,\phi) = [\hat{A}_\alpha p^{(n)}](\theta,\phi)\,,
\end{equation}
where the operator $\hat{A}_\alpha$ averages its operand over all
points that are one angular step $\alpha$ away from the target point
$(\theta,\phi)$.  For example, if the initial distribution is a
delta-function $p_\delta(\theta,\phi) = \delta(\theta)$ at the north
pole, after one step the distribution is the uniform average over
the ring of points $\{(\alpha,\phi)\,|\,\phi\in[0,2\pi)\}$.

Because the average of a sum is the sum of the averages,
$\hat{A}_\alpha$ is linear and can be diagonalized. Assuming the
step size $\alpha$ is fixed, the eigenfunctions are the spherical
harmonics, with eigenvalues given by the Legendre
polynomials~\cite{RU},
\begin{equation}
  [\hat{A}_\alpha Y_{\ell m}](\theta,\phi) = P_\ell(\cos\alpha)
  Y_{\ell m}(\theta,\phi)\,.
\end{equation}

If we expand an initial azimuthally-symmetric distribution
$p^{(0)}(\theta)$, expressed in the basis of Legendre polynomials as
\begin{equation}
  p^{(0)}(\theta) = \sum_{\ell=0}^\infty c_\ell^{(0)} P_\ell(\cos\theta)\,,
\end{equation}
then the distribution after an $n$-step random walk is given by
\begin{equation}
  \label{eq:ClassicalDistributionWalking}
  p^{(n)}(\theta) = \sum_{\ell=0}^\infty c_\ell^{(0)}
  P_\ell(\cos\theta) [P_\ell(\cos\alpha)]^n\,.
\end{equation}
The coefficients $c_\ell^{(n)}$ for the distribution at any timestep
$n$ are therefore given in terms of the initial coefficients as
\begin{equation}
  \label{eq:CoefficientEvolution}
  c_\ell^{(n)} = c_\ell^{(0)}[P_\ell(\cos\alpha)]^n \,.
\end{equation}
This provides us with the classical evolution of the DRF.

\subsection{Fitting the quantum degradation to a random walk}

We now consider how the distribution $p^{(0)}(\theta)$ of
Eq.~\eqref{eq:InitialClassicalDistribution}, undergoing a random
walk on the sphere with angular step size $\alpha$, serves as a DRF
according to the classical average measurement fidelity as a figure
of merit.

For the distribution $p^{(0)}(\theta)$ of
Eq.~\eqref{eq:InitialClassicalDistribution} expressed as an
expansion in terms of Legendre polynomials as in
Eq.~\eqref{eq:DistnSphericalHarmonics}, the first two coefficients
are found to be
\begin{equation}\label{eq:c0c1}
    c_0^{(0)} = 1\,,\quad
    c_1^{(0)} = \frac{6j}{2j+1} \,.
\end{equation}
The first two coefficients after $n$ steps of the random walk are
given by Eq.~\eqref{eq:CoefficientEvolution} to be
\begin{equation}\label{eq:c0c1(n)}
    c_0^{(n)} = 1 \,,\quad
    c_1^{(n)} = \frac{6j}{2j+1}\cos^n\alpha \,.
\end{equation}
The classical average measurement fidelity after $n$ steps, using
Eq.~\eqref{eq:cfidc0c1}, is thus
\begin{equation}
  F_C(n) =  \frac{1}{2}+\frac{j}{2j+1}\cos^n\alpha \,.
\end{equation}
We note that this fidelity is precisely equal to that of the
fully-quantum case, given by Eq.~\eqref{eq:PsuccessSU(2)CS}, if the
step size $\alpha$ of the random walk is given by
\begin{equation}\label{eq:varystep}
  \cos\alpha = 1-\frac{2}{(2j+1)^2}\,.
\end{equation}
As argued above, the dependence on $j$ of this step size appears
reasonable, satisfying $\alpha \simeq 1/j$ for large $j$.
However, we emphasize that this parameter is obtained from the full
quantum theory, as it is the fundamental quantum back-action that
determines its magnitude.

Given that the classical model reproduces the average
measurement fidelity of the quantum model \emph{exactly}, i.e., to
all orders in $1/j$ when a step size given by
Eq.~\eqref{eq:varystep} is used, it is natural to question whether
the full quantum evolution of the DRF is captured by our
semi-classical model. Specifically, one might naively suggest that
an initial SU(2) coherent state $|j,j\rangle$ undergoing a random
walk on the sphere with step size~\eqref{eq:varystep} would yield a
mixed quantum state of the DRF that provides a solution to the
quantum evolution of Eq.~\eqref{eq:RFdecoherence}.  Surprisingly,
this is not the case: we have checked numerically that solutions to
Eq.~\eqref{eq:RFdecoherence} \emph{cannot} be expressed as convex
combinations of SU(2) coherent states.  Thus, we emphasize that,
while the degradation (as quantified by the average measurement
fidelity) is captured precisely by our semi-classical model, the
full evolution of the quantum DRF is not.

\section{Conclusions}
\label{sec:Conclusions}

We have shown here that, given a suitable initial classical
distribution for the misalignment of an ideal DRF and a reasonable
semi-classical measurement theory that incorporates back-action, the
exact expression of the decreasing measurement fidelity
describing the degradation of a quantum DRF is reproduced by a
classical random walk using a single fitted parameter -- the
step size of the random walk.

As the step size was fitted to the back-action from the full
quantum result, a semi-classical argument on its own does not
precisely explain the quadratic scaling of a reference frame's
longevity~\cite{BRST}.  However, one can provide an heuristic
argument for quadratic scaling.  Note that, at least in the case of
a directional reference frame, it is the $1/j$ behavior of the angle
$\alpha$, that leads to quadratic scaling. The simple classical
picture of adding a small vector randomly to a vector of length $j$
shows that the angle $\alpha$ between the resulting sum of vectors
and the original vector will scale as $1/j$, leading to a quadratic
scaling in longevity.

We emphasize that our results are applicable only to the case where
the measured systems have zero mean polarization relative to the
DRF. A natural extension of this work would be to incorporate into
this analysis the linear drift of the DRF that occurs when this
polarization is non-zero. Such a drift would be straightforward to
incorporate into the classical random walk.  The key challenge
appears to be the derivation of an analytical expression for
Eq.~\eqref{eq:PsuccessSU(2)CS} for the case of non-zero bias.

We note that, in modeling the quantum DRF with a semi-classical
model, the classical analogue of the quantum state of the frame was
a \emph{probability distribution} over possible classical
configurations, i.e., a state of \emph{knowledge} about the
orientation of a classical DRF.  This was true even for the initial
pure quantum state $|j,j\rangle$.  Viewing quantum states as states
of knowledge, analogous to classical probability distributions, is
the principle tenet of the epistemic view of quantum
states~\cite{Fuc02,Spe04}.  Our results provide further potential
for advancing this research program.  First, our results demonstrate
that, for the specific case of measurements relative to a
classical DRF, the degradation of the measurement fidelity is
described by a hidden-variable model.  In addition, our results
highlight the importance of incorporating a form of back-action into
the measurement theory within such a hidden-variable model.  The
existence (and necessity) of back-action associated with measurement
is also a feature of ``toy'' models based on the principles of the
epistemic view that reproduce quantum-like phenomena~\cite{Spe04}.

\section{Acknowledgments}

We wish Peter Knight well on this special occasion and appreciate
his support and guidance throughout the years.

The authors gratefully acknowledge J-C Boileau, David Poulin, and
Robert Spekkens for helpful discussions.  SDB acknowledges support
from the Australian Research Council (ARC).  TR acknowledges support
from the Engineering and Physical Sciences Research Council of the
United Kingdom, and a University of Sydney Short-Term Visiting
Fellowship. BCS acknowledges support from Alberta's Informatics
Circle of Research Excellence (iCORE), ARC, and the Canadian
Institute for Advanced Research. PST acknowledges support from an
iCORE Alberta Ingenuity Fund Fellowship, and from the Network of
Centres of Excellence for the Mathematics of Information Technology
and Complex systems.

\end{document}